%% file: tsqt6.tex
\documentclass{article}
\usepackage{bezier,emlines2}
\advance\textheight by -\headsep\advance\textheight by -\headheight
\begin{document}

\title{Time-Symmetrized Quantum Theory, Counterfactuals,
and `Advanced Action'}
\author{ R.E. Kastner\thanks{rkastner@wam.umd.edu}}
\date{July 1, 1998}
\maketitle
\begin{center}
{\small \em Department of Philosophy \\
University of Maryland \\
College Park, MD 20742 USA. \\}
\end{center}
\vspace{.1cm}
\begin{abstract}
		
Recent authors have raised objections to the counterfactual
 interpretation of the Aharonov-Bergmann-Lebowitz (ABL) 
rule of time symmetrized quantum theory (TSQT).
  I distinguish between two different readings of the ABL rule,
 counterfactual and non-counterfactual, and confirm that 
TSQT advocate L. Vaidman is employing the counterfactual
 reading to which these authors object.  Vaidman  has responded 
to the objections by proposing a new kind of time-symmetrized 
counterfactual, which he has defined in two different ways. 
 It is argued that neither definition succeeds in overcoming
 the objections, except in a limited special case previously 
noted by Cohen and Hiley. In addition, a connection is made
 between TSQT and Price's concept of `advanced action',
 which further supports the special case discussed.
\end{abstract}
\large
{\bf 1. Introduction.~~ }
\vskip .2cm

Time-Symmetrized Quantum Theory (hereinafter TSQT) was first 
introduced by Aharonov, Bergmann and Lebowitz (1964), and most
recently defended by Vaidman (1996a, 1997a,b).  The basic claim 
of TSQT (actually a novel {\it interpretation\/}
of quantum theory) is that a fundamental time symmetry applies 
to the interval between two ideal measurements. According to 
TSQT, the maximally specified quantum state of a system between 
two measurements occurring at times $t_1$ and $t_2$
contains information based not only on the initial, or 
pre-selection measurement (yielding the state
 $|\psi_1(t_1)\rangle)$ but also on the final, or post-selection
 measurement (yielding the state $|\psi_2(t_2)\rangle)$. 
Ensembles of systems identified in this way are referred
to as ``pre- and post-selected ensembles.'' Aharonov and 
Vaidman (1991) proposed that such ensembles be  represented 
by a two-state vector, or ``generalized state'' 
$\Psi(t)=\langle \psi_2(t)||\psi_1(t)\rangle$, where 
$|\psi_1(t)\rangle= U(t, t_1) |\psi_1(t_1)\rangle$, and 
 $\langle \psi_2(t)|=  \langle \psi_2(t_2)| U(t_2, t)$. 
The propagator $U(t_2, t)$ represents a time-reversed evolution.

The novel philosophical claim of TSQT, as compared to most of
 the traditional interpretations of quantum mechanics--
including interpretations as diverse as the Copenhagen 
interpretation, many worlds theories, and hidden variables
theories such as Bohm's theory--is that the ontological state
of a quantum system has a causal dependence not only on past 
measurements, but on measurements to be made in its immediate 
future.  In this respect, TSQT advocates have much in
common with the philosophy of time of Price (1996), who argues
that a fundamental time symmetry holds at the quantum level
and that the world exhibits what he calls ``advanced action,''
essentially backwards causality. Indeed, Vaidman
himself notes that the TSQT formalism ``is not too far from
 the spirit of [Price's] `advanced action'....''
 (1996, footnote 6).  This connection between advanced 
action and TSQT will be discussed in more detail in Section 4.

Sections 2 and 3 of the paper will focus on the claim of Sharp
and Shanks (1993)  to have refuted TSQT. It will be argued 
that, despite Vaidman's arguments to the contrary 
(1996a, 1997a,b),  these authors have successfully refuted 
a broad claim of the theory: namely, the assertion that a 
probability rule derived by Aharonov, Bergmann and Lebowitz 
in their (1964)--hereinafter `the ABL rule'-- can be
universally applied to counterfactual measurements at times 
between the pre- and post-selection measurements.  
Section 4 discusses a special case, also noted in a slightly
different context by Cohen (1995) and Cohen and Hiley (1996),
in which the Sharp and Shanks criticism fails and the 
counterfactual usage of the ABL rule is valid. 
\vskip .2cm

{\bf  2. TSQT and counterfactual claims.~~}

\vskip .2cm	
In this section, I first review the ABL rule for calculating 
the probability of a given outcome of a measurement occurring
between pre- and post-selection measurements. I then discuss
proposed interpretations of the ABL rule.  I distinguish 
between two readings, non-counterfactual and counterfactual. 
The non-counterfactual version is trivially correct; I argue
that Vaidman is in fact employing the counterfactual version. 
Subsequently, I review the Sharp and Shanks  proof that the 
counterfactual use of the ABL rule gives results inconsistent
with quantum mechanics, and Vaidman's counterarguments. I show 
that both of his arguments fail:  one (henceforth called 
``counterargument I'') relies on the possibility of allowing 
for a counterfactual with true antecedent which I argue is 
irrelevant to the counterfactual usage under debate, and the
other (henceforth called ``counterargument II'') fails to solve
(or even to take into account) the cotenability problem which
is at the root of the general inapplicability of a
counterfactual usage of the ABL rule. 

The original ABL paper (1964) derived an expression for the 
probability of an outcome $c_j$ of a measurement of a
non-degenerate observable $C$ with eigenvalues $\{c_i\}$ at a
time $t$ between pre- and post-selection measurements at times 
$t_1$ and $t_2 \ 
 (t_1 < t < t_2)$. This expression, subsequently
known as the ABL rule, is an uncontroversial result of
quantum mechanics, applied to the situation in which the 
intervening $C$ measurement has actually occurred.
In the case of zero Hamiltonian, which is sufficient
for the purposes of this paper, 
the rule states that the probability of  outcome $c_j$ in the 
case of a preselection
for the state $|a\rangle$ and a post-selection for the state 
$|b\rangle$ is given by:

	$$ P_{ABL}(c_j|a,b) = { |\langle b|c_j\rangle|^2 
|\langle c_j|a\rangle |^2 \over{\sum_i |\langle b|c_i
\rangle |^2 |\langle c_i|a\rangle |^2}}   \eqno(1)$$

\vskip .2cm
	More recently, some authors (Albert, Aharonov and 
D'Amato, 1985; hereinafter AAD) began to make explicit use 
of a counterfactual interpretation of the ABL rule. 
Rather than just using the rule to retrodict the probability
 of an outcome $c_j$ for a measurement of $C$ which was actually 
carried out, AAD argued that the rule could be interpreted 
as ``the probability....that a measurement of some complete 
set of observables $C$ within [the interval $(t_1, t_2)$],
 if it were carried out, would find that $C=c_j$''
 (1985, p. 5). 	

For clarity, I shall distinguish between
 two different readings of the ABL rule which, at first 
sight, both appear to be counterfactual. The first,  
non-counterfactual, reading asserts for a suitably pre- 
and post-selected system  and  noncommuting observables 
$X,Y$,....:

\hangindent 3pc \hangafter -3	``If a measurement of $X$ was
 performed at time $t$, then the probability that result $x$ 
was obtained is $p(x)$; and if a measurement of $Y$ was 
performed at time $t$, then the probability that result $y$ 
was obtained is $p(y)$, and if....''
\vskip .2cm

	Such a non-counterfactual reading is discussed in
 Mermin (1997).  It is essentially a conjunction of material
 conditionals,  only one of which is non-vacuously true,
 because only one of the antecedents can be true at time $t$. 

A second  reading of the ABL rule,  the {\it bona fide}
 counterfactual  usage, asserts that for {\it any} observable
$X$, the ABL probability gives the correct probability that the property
associated with eigenvalue $x$ is possessed by the system.
Thus, the counterfactual usage makes
the following counterfactual prediction for a given 
pre- and 
post-selected system:
\vskip .2cm
\hangindent 3pc \hangafter -2
	``If I had performed a measurement of observable $X$ 
on the system at time $t$, then the probability of the result
 $x$ would have been $p(x)$, where $X$ is any observable.''
\vskip .2cm

This is the type of claim being made in AAD (1985) as noted 
above. The counterfactual reading is also used, for example, 
by Aharonov and Vaidman (1991) to argue that a single 
particle with suitably pre- and post-selected states can be 
found with certainty in N separate boxes, as well as in other
 examples of `curious' quantum effects involving pre- and 
post-selected ensembles.\footnote{\normalsize The argument crucially
depends on the assumption that the given particle has legitimate
counterfactual ABL probabilities for different possible
intervening measurements.}

	The possibility of the above two distinct readings of
 the ABL rule (1) arises because of quantifier ambiguity in 
the expression $P_{ABL}(c_j\vert a,b)$.
  (In the following we omit the ABL subscript for brevity.)
 As originally derived, this expression is actually shorthand
 for $P(c_j\vert a,b;C)$, where $C$ is the observable 
actually measured at time $t,$  $t_1<t<t_2$, in the
 selection of a particular system with generalized state
 $\langle b\Vert a\rangle$. Thus the explicitly characterized
 ABL rule is:

$$ P(c_j|a,b;C) = { |\langle b|c_j\rangle|^2 
|\langle c_j|a\rangle |^2 \over{\sum_i |\langle b|c_i
\rangle |^2 |\langle c_i|a\rangle |^2}}   \eqno(1')$$

	Omission of the parameter $C$ as in (1) creates an
 ambiguity in the variable $c_j$: namely, whether or not it
represents values of 
the observable actually measured. To make this ambiguity more
 evident, we will from now on use the parameter $C$ to 
indicate the observable that is
 actually measured at time $t$, and replace the variable 
$c_j$ with the more general variable $x_j$; i.e.,
$x_j$ might represent a value from some observable $O$ rather
than $C$. Thus the
 explicitly ambiguous form of the ABL rule can be written as:

	   $$ P(x_j|a,b;C) = { |\langle b|x_j\rangle|^2 
|\langle x_j|a\rangle |^2 \over{\sum_i |\langle b|x_i
\rangle |^2 |\langle x_i|a\rangle |^2}}   \eqno(1'')$$

	An equation like $(1'')$ is always understood as 
implicitly quantified. The two different readings correspond
 to two different quantifications. Specifically, 
the non-counterfactual reading asserts that equation $(1'')$
 holds for all $a,b \in {\cal S}$, all $C \in {\cal O}$,
 and all $x_j \in {\cal R}(C)$, where  $\cal S$  is the set 
of pure states, $\cal O$ is the set of observables, and 
${\cal R}(C)$  is the range of values of $C$. (The possibility 
that no measurement was performed at time $t$ is covered if 
we include the identity  $I$ in the set ${\cal O}$.)

	The counterfactual reading  differs from the above
 in that equation $(1'')$ is asserted to hold for all $a,b \in 
{\cal S}$, all $C \in {\cal O}$,
 and all $x_j \in {\cal R}(O)$, for {\it all} $O \in {\cal O}$,
  including the 
cases where $O \ne C$.  Thus  the counterfactual reading  
allows $x_j$ to vary over the values of observables that were
 not actually measured at time $t$--and, of course, there is
 a fact of the matter as to whether or not a measurement was
 performed at time $t$, and if so, which observable was measured.

	Vaidman (1996b) effectively employs the counterfactual
 reading by arguing in his discussion of a single particle
 inside three boxes that the outcomes receiving probability 
one according to the ABL rule correspond to `elements of
 reality' for that particle:
\vskip .2cm

	``Elements of reality in the pre- and post-selected 
situations might be very peculiar. One such example is a 
single particle inside three boxes A, B, and C, with two 
elements of reality: `the particle is in box A' and `the 
particle is in box B' ...If in the intermediate time it was 
searched for in box A, it has to be there with probability 
one, and if, instead, it was searched for in box B, it has
 to be found there too with probability one.''\footnote
{\normalsize Vaidman (1996b), p. 900. Vaidman uses box numbers instead
of letters; I modify this for uniformity of notation.}
\vskip .2cm
	This usage clearly implies that the properties of
 being in box A or being in box B are considered as
 possessed by the same pre- and post-selected
 particle. Since a measurement of some observable $O$ was in 
fact performed at time $t$ between the pre- and post-
selection of the particle (where $O$ is either the observable 
associated
 with the particle being in box A, or the observable
associated with the particle being in box B), or no
 measurement was performed (in which case $O=I$), it is
 clear that Vaidman is allowing $x_j$ to vary over the values
 of observables  $O' \ne O$ that were not actually measured 
between the pre- and post-selection of that particular 
particle. \footnote{\normalsize The phrase, ``If, instead, it was
searched for in box B'' directly trades on the ambiguity
concerning whether the $B$ measurement was {\it actually}
performed in the selection of that particle. If this is the
 case,  the left hand side of the ABL rule as formulated
 in $(1'')$ 
becomes $P(b\vert \psi_1,\psi_2;B)$ (with $\psi_1$
 and $\psi_2$ being the pre- and post-selected states,
respectively, and $b$ the eigenvalue corresponding to
 finding the particle in box B): the
non-counterfactual reading. But if one can use the
 non-counterfactual reading for the $B$
`element of reality,' then one is committed to the $B$
 measurement being the one which was
actually performed in the selection of the given particle. 
Then the proposition concerning
the $A$ `element of reality' uses the counterfactual reading,
 since the left hand side of $(1'')$
in this case would be $P(a\vert \psi_1,\psi_2;B)$,
$a$ being the eigenvalue corresponding to finding
the particle in box A; and $a$ is not in the range of 
values of $B$.
 The analogous argument applies to
the case in which A is the observable actually measured.}

	However,  Sharp and Shanks (1993, hereinafter S\&S)
 present a proof that the counterfactual interpretation of
 the ABL rule leads to predictions incompatible with quantum
 mechanics.  They consider an ensemble of spin-$1\over2$
 particles  prepared at time $t_1$ in the state 
$\vert a_1\rangle$
 (read as `spin up along direction {\bf a}'). They then
 assume that this ensemble is subjected to a final post-
selection spin measurement at time $t_2$ along direction
{\bf b} (i.e., the observable $\sigma_b$ is measured).
This measurement yields a mixture, which we shall call M,
 consisting of two subensembles $E_i, i=1,2$, described by the 
two-state vectors $\langle b_i\Vert a_1\rangle$,
 where the subscript '2' denotes spin down.  The weight of 
each subensemble $E_i$ is given by
 $\vert \langle b_i\vert a_1\rangle \vert ^2$.
 (See Figure 1.)  
\hrule
\input{fig1}
\newline \small
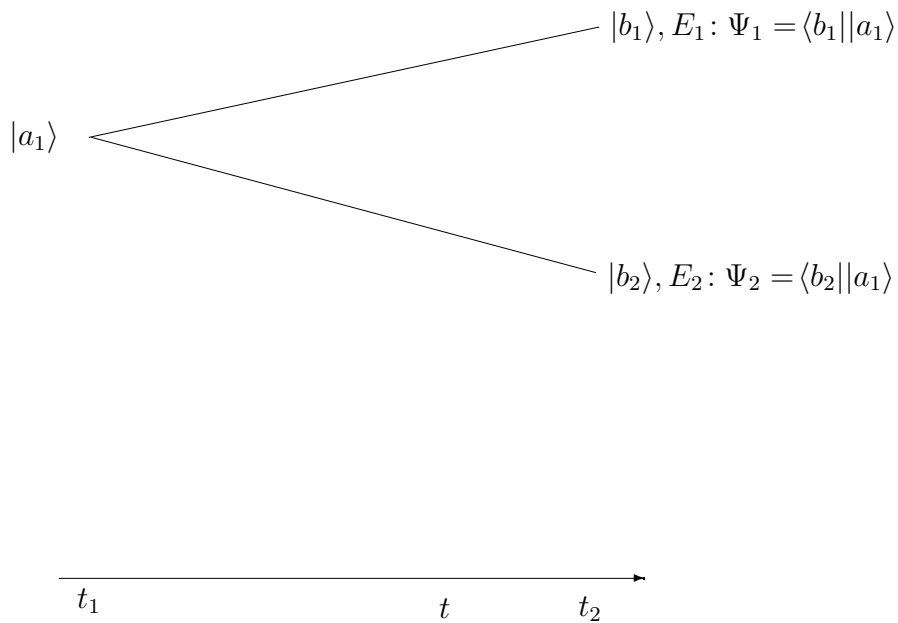
Figure 1. No intermediate measurement is made.
 Two ensembles result: $E_1$, with two-state vector 
$\Psi_1=\langle b_1||a_1\rangle$; and $E_2$, with two-state
 vector $\Psi_2=\langle b_2||a_1\rangle$.
\hrule
\vskip .1cm \large
	Now they consider each subensemble individually, 
asking the counterfactual question: If we had measured the
 spin of these particles along direction {\bf c}
 (i.e., observable $\sigma_c$)  at a time $t$ between \  $t_1$ \ and \ $t_2$, what
 would have been the probability for outcome $c_1$? They use
the ABL rule to calculate the probability of  outcome $c_1$
 for each subensemble $E_i$ for such a counterfactual
 measurement. They then show that the total probability of 
outcome $c_1$ derived from the above calculation, taking into
 account  the weights of the two subensembles $E_i$, in
 general disagrees with the quantum mechanical probability,
 which is given simply by
 $\vert \langle c_i\vert a_1\rangle \vert ^2$.

	The result obtained by S\&S depends on the 
assumption that the intervening $\sigma_c$ measurement has
 definitely not occurred. In  counterargument I,
 Vaidman (1996a) argues that it is permissible for his
 purposes to assume that the $\sigma_c$ measurement actually 
occurred because he wants to allow for the possibility of a
 counterfactual that is not necessarily `contrary to fact'.
 While Lewis (1973, p. 26) has argued cogently that one can
 have a counterfactual with a true antecedent,  such an 
argument is of no use here because Lewis' counterfactual with
 true antecedent simply  reduces to a material conditional.
 This brings us back to the ordinary non-counterfactual 
reading of the ABL rule, which is not under dispute.

	In counterargument II, Vaidman (1996a, 1997a) claims
 that the S\&S calculation is not the right one. He shows that
 if one includes the dependence of the final measurement on
 the intervening measurement, then the resulting $P(c_1|a_1)$
  is in fact equal to $\vert \langle c_1\vert a_1\rangle \vert ^2$.
 
 However, what this `correction' amounts to is assuming
 that the putative counterfactual measurement has actually
 occurred--just as in counterargument I--since the ensembles
  have been redefined to  include the dependence on the 
intervening measurement.  Thus, with the ensembles redefined
 in this way, it appears we are no longer considering a 
counterfactual conditional of the form
\def\cf{\raisebox{3 pt}{\fbox{}}\!\! \to}

$$ P \ \cf Q\eqno(2)$$ 

(``If it had been the case that P, then it would have been
 the case that Q'')
where $P$= ``observable $C$ is measured''
 and  $Q$ = ``the probability of outcome $c_1$ is as given
 by the ABL rule (applied to systems appropriately pre- and 
post-selected according
to the procedure of S\&S)'', 
but instead the material conditional

	 $$P  \to  Q.  \eqno(3)$$
						
\def\ba{$\langle b_1\Vert a_1\rangle$}
\def\bba{$\langle b_2\Vert a_1\rangle$}
\def\mprime{M\raisebox{1 pt}{$^\prime$}}

	The basic conceptual problem is the following: 
in considering a counterfactual measurement of the spin along
 {\bf c} (observable $\sigma_c$), we must take into account 
all the effects of that measurement on the system. 
Measurement of the observable $\sigma_c$ results in a change
 in the mixture M of post-selected ensembles $E_i$
 described by the two-state vectors \ba \ and \bba :
 were we to make the intervening measurement, we would not
 have the mixture M but rather the mixture \mprime  consisting of
 the four subensembles $E'_{jk}$ with weights
\def\bkcj{\vert \langle b_k\vert c_j\rangle \vert^2}
\def\cja1{\vert \langle c_j\vert a_1\rangle \vert^2}
$\bkcj\!\cja1$, where $j, k =1,2.$  (See Figure 2.)
\vskip .1cm
\hrule
\input{fig2}
\small\nopagebreak\vskip 1pt \noindent
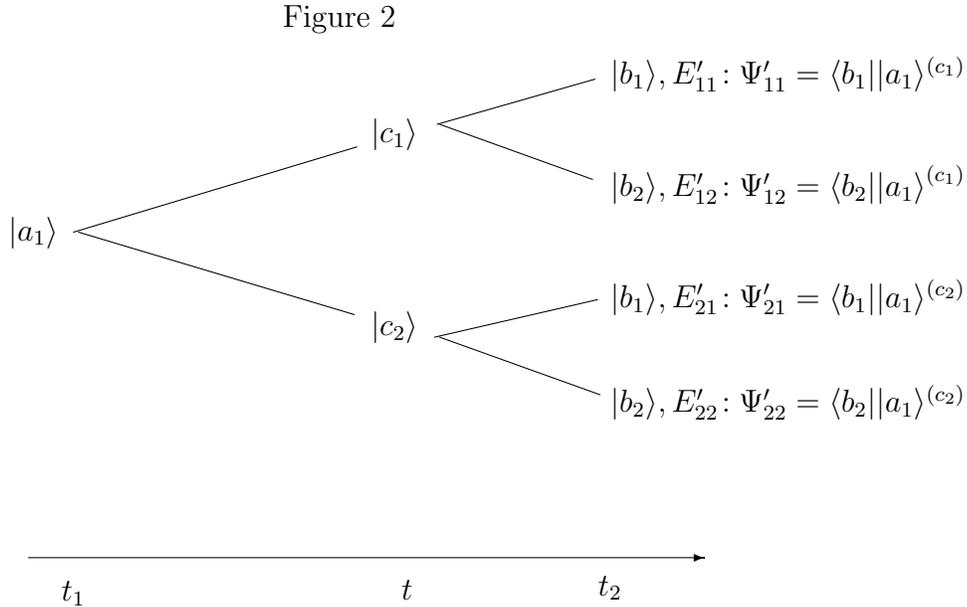
Figure 2: Intermediate
 measurement of $\sigma_c$. Four sub-ensembles result:
$E'_{jk}, j,k=1,2$, with associated two-state vectors 
$\Psi'_{jk}= \langle b_k||a_1\rangle^{(c_j)}$. The superscript
$(c_j)$ indicates that the given sub-ensemble was obtained
from systems yielding outcome $c_j$ at time $t$. We also 
define ensembles $\eta_k = \sum_j E'_{jk}$ to compare
the statistical weights of those ensembles having 
final outcome $b_k$ with those of the $E_i$, where
$i=k$.
\hrule
\vskip .2cm
\large This dependence of  the actual mixture, M or \mprime,  obtaining 
at  time $t_2$ on whether or not an intervening measurement is 
performed gives rise to a crucial difficulty which destroys 
the validity of the counterfactual claim.  In order to 
clarify this point, and to lay the groundwork for a detailed
 critique of Vaidman's counterargument II, we need some ideas
 from philosophical theories of counterfactuals.  
	
	The intuitive way to formulate truth-conditions 
for counterfactual statements is probably best exemplified
 by  Goodman's classic essay, ``The Problem of Counterfactual
Conditionals'' (1947). Goodman's approach, now commonly
 referred to as the 
metalinguistic account of counterfactuals, proposes that a 
counterfactual such as			
$P \ \cf Q$ is true if and only if 
its antecedent $P$ implies (i.e., semantically entails) the
 consequent $Q$ when $P$ is conjoined with suitable
 background conditions $S$ and laws of nature $L$:
\vskip .2cm
 \hfil $P\ \cf Q$ is true iff 
 $(P  \& S \& L) \models  Q$.\hskip 5cm (4)
				 
\vskip .2cm
	In addition to (4), a further condition is
 necessary to specify what sorts of facts can be included 
in the collection of `suitable background conditions' $S$.
  Some restrictions are no doubt required, because certain
 propositions--for instance $\neg P$--result in the
 conjunction in (4) being a contradiction, in which case 
$P\ \cf Q$ is vacuously true 
(i.e., true for any $Q$).  Metalinguistic  accounts describe
 the restriction on membership in $S$ in terms of 
`cotenability': the notion that the holding of any condition 
belonging to $S$ must be independent of the truth value of 
$P$.  Thus condition (4) must be augmented by a further
 requirement :
\vskip .2cm
  \hfil The background conditions asserted by $S$ must
 be cotenable with $P$. \hfil (4a).         
\vskip .2cm
 For future reference, let us refer to the condition defined
 by (4) together with (4a) as `condition $\Gamma$'.

	However (as discussed by Goodman, and more recently
 in Horwich 1988, Chapter 10), the cotenability notion can 
only be defined within the metalinguistic account in terms 
of the counterfactual $P\ \cf S$;\footnote{\normalsize More precisely,
the condition for cotenability of $S$ with $P$ in the metalinguistic
account is the 
conjunction $(\neg P \& S)\ \& \ (P\ \cf S)$.}
 so the metalinguistic account faces a circularity problem
 in giving a satisfactory account of cotenability.
 Fortunately, one has recourse to  more successful theories 
of counterfactuals in order to give a rigorous account of
 cotenability (for example, Lewis, 1973; and Stalnaker, 
1968). While these theories are based on the notion of 
possible worlds, they can also be expressed in terms of
 cotenability, as Lewis (1973, p. 57) has shown. The failure
 of the counterfactual interpretation of the ABL rule can 
most easily be analyzed in terms of a failure of cotenability
 between the statement of background conditions, $S$, and 
the antecedent $P$. Therefore, for convenience, the analysis
  of truth-conditions for counterfactuals will first be
 discussed in terms of the metalinguistic account's
 cotenability notion, and later related to the possible 
worlds formulation.

	Returning now to  the discussion following equation
 (3), the counterfactual usage of the ABL rule as studied in
 the S\&S example faces a difficulty in that the mixture, 
M or \mprime, obtaining at time $t_2$ depends on whether or not an
 intervening measurement has been performed at time $t$
 (compare figures 1 and 2). Since the mixture must enter
 into the  counterfactual calculation in the form of the 
weights of each of the component subensembles, with their 
associated two-state vectors used as inputs in the ABL
 formula (1),  the characterization of the mixture obtaining
 at $t_2$ must be included in the background condition 
statement  $S$ in (4).  Thus, the statement of background conditions
 $S$ holding when $\neg P$ holds becomes {\it false} when 
 $P$ holds and is therefore not cotenable with $P$ (see footnote 4).
 By this criterion alone, it is clear that
 the counterfactual usage of the ABL rule fails because
 condition $\Gamma$ is not met: (4a) fails.

The S\&S proof goes further than this by demonstrating that
 the counterfactual ABL probability based on the statement 
of the actual background condition holding when $P$
 is false (i.e., when $C$ is not measured and mixture M obtains)
is incorrect.
 (Note that had this background condition statement been
 cotenable with $P$, the same procedure would yield a 
correct counterfactual probability. This is the case in the 
usual quantum mechanical counterfactual calculation of the
 probability of the outcome of a measurement at time $t$
 given the quantum state at time $t'$.  In this case the 
background condition is simply the quantum state at time 
$t'$,  which does not depend on whether or not a subsequent 
measurement is performed).  
 
	We now describe in detail the steps employed in the 
S\&S proof. First, consider the overall background
 conditions $S$ obtaining when $P$ is false.  $S$
 consists of the statement of contingent initial conditions 
$S_1$ concerning events at time $t_1$, as well as a
 statement concerning events at time $t_2$, which we shall
 call $S_2$. Thus, $S = S_1 \& S_2$.
  In the S\&S proof, $S_1$ corresponds to  the initial 
preselected ensemble of particles in state $|a_1\rangle$.
 $S_2$ states
 that mixture M obtains at time $t_2$. It is clear that $S_2$ 
follows from $S_1$ when $P$ does not hold; i.e., from
 the laws of quantum theory asserted by $L$ we have:
\vskip .2cm
$$S_1 \& L \& \neg P \models  S_2.\eqno(5)$$
\vskip .2cm
\noindent S\&S then consider a hypothetical case which differs
 from the actual one only in that the antecedent $P$ is
 true; i.e., $C$ is measured at $t$. 
They derive the consequence  $R$ of the conjunction
 of the background conditions $S=S_1 \& S_2$ , relevant
 quantum mechanical laws $L$, and  $P$:
\vskip .2cm
	 $$  S \& L \& P \models  R.\eqno(6)$$ 
\vskip .2cm
	 $R$ states: ``The probability of outcome $c_1$
 of an intervening measurement of observable $C$ is as 
given by the ABL rule and the rule of total probability 
applied to systems in the mixture M.''  Thus, $R$ is 
equivalent to $Q$ in (2), since the systems referred to 
in $Q$ are the systems obtaining when
$P$ does not hold, which must be those in mixture M.
	The explicit calculation corresponding to (6) is:
\vskip .2cm
 $$ P(c_1|a_1) = P_{ABL}(c_1|a_1\&b_1)P(b_1|a_1)
 + P_{ABL}(c_1|a_1\&b_2)P(b_2|a_1),\eqno(7)$$ 
\vskip .2cm \noindent											
where the left hand side of (7), $P(c_1|a_1)$, is the
 probability referred to in the proposition $R$.
 On the right hand side, the quantities 
$P_{ABL}(c_1|a_1\&b_k), k=1,2$ 
 correspond to the component of $L$
 asserting that the ABL rule gives the correct quantum 
mechanical probability for the given outcome of an
 intervening measurement for systems with pre- and 
post-selected states $\langle b_k\|a_1\rangle$.
 The weighting factors
$P(b_k|a_1), k=1,2$ correspond to the component of $L$ 
stating that the rule of total probability holds, 
together with background condition statement $S=S_1\&S_2$,
 since the weights are those obtained from ensembles $E_i,
 i=1,2$. In terms of quantum theory, the quantities
 $P(b_k|a_1)$ are given by
\vskip .2cm
	$$P(b_k|a_1) = |\langle b_k|a_1\rangle |^2,\eqno(8)$$
\vskip .2cm
which of course assumes that no intermediate measurement of 
$\sigma_c$ has been performed.

	However, note that the left hand side of (6) is false
 because it is not possible (in general)\footnote{\normalsize The exception
is when the post-selected systems referred to in $S$
happen to be the same whether or not $P$ is true; for now
we neglect these cases, which will be discussed in \S 4.}
 for both $P$ and $S_2$  to be true,  given $S_1$
 and $L$ (this being the failure of cotenability discussed
 earlier).  Therefore, the consequent $R (=Q)$  may well be 
false. 

	The S\&S strategy is to show that $R$ is indeed 
false, by comparing $R$ to the true consequent $Q'$
 which results in the case in which $P$ holds and the
 problematic statement of background conditions $S$
 has been replaced so as to ensure that the conjunction 
appearing on the left hand side of (6) is true.  Thus, S\&S
 next consider what would actually occur given the truth of 
$P$. There are two equivalent ways to do this. What S\&S
 do is simply to drop the inconsistent background condition
 statement $S_2$, and derive the entailment $Q'$ of the 
conjunction of $S_1$ , $P$ and  $L$.
 Thus they perform the derivation:
\vskip .2cm
	$$S_1 \& P \& L  \models  Q'.\eqno(9)$$	
\vskip .2cm					(9)
The proposition $Q'$ states ``The quantum mechanical 
probability for outcome $c_1$ for the given ensemble of
 particles (in this case, the pre-selected only ensemble)
 is as given by the usual quantum mechanical conditional 
probability.''  Thus the explicit calculation corresponding 
to (9) is:
\vskip .2cm
	$$P_{QM}(c_1|a_1) = |\langle c_1|a_1\rangle|^2
\eqno(10)$$ \vskip .2cm
Since the left hand side of (9) is now true,
  the consequent $Q'$ must be true. 

However, it may not be
 readily apparent that $Q'$  really is 
the appropriate proposition to compare with $R$, since it does 
not directly incorporate the ABL rule. In order to show that
 $Q'$ is the correct proposition to compare with $R$,
 we now
 consider the second, equivalent way of deriving the 
consequence of the conjunction of the correct background
 conditions, $L$, and $P$. This consists of deriving the analog
 of $R$ in (6) for the case in which the background conditions
 are those obtaining when $P$ is true. These correct background
 conditions, which we shall call $S'$, consist of the
 conjunction of $S_1$ and $S_2'$, where $S_2'$  states that
 subsequent to the intervening measurement $P$ we have mixture 
\mprime consisting of the four 
subensembles $E'_{jk}$ and their associated weights. 
	
	In order to facilitate the comparison of $S_2$ with
 $S_2'$, we define the ensembles 
$\eta_k = \sum_j E'_{jk}, j,k=1,2$, where $\eta_k$ is  the 
ensemble composed of 
all systems found in state $|b_k\rangle$ at time $t_2$
 for the case in which $P$ holds (see Figure 2). 
	Now, in direct analogy with (6), we can  derive the
 consequence $R'$ of the conjunction of $S'$ with $L$ and 
$P$ :
\vskip .2cm
$$	  S' \& L \&  P \models R'.\eqno(6')$$
\vskip .2cm
Note, just as in (9),  that the left hand side of (6')
 is true, so the consequence $R'$ must be true. The 
proposition $R'$ states:  ``The probability for outcome
 $c_1$ of an intervening measurement of observable $C$
 is as given by the ABL rule and the rule of total
probability applied to systems in the ensembles 
$\eta_k = E'_{1k} + E'_{2k} , k=1,2$,
 resulting when $P$ holds.''
 
	The explicit calculation corresponding to (6') is:
\vskip .2cm	
	$$P_{QM}(c_1|a_1) =  P_{ABL}(c_1|a_1\&b_1)
 P_{\sigma_c}(b_1|a_1)	+ P_{ABL}(c_1|a_1\&b_2)
 P_{\sigma_c}(b_2|a_1)\eqno(12)$$
 \vskip .2cm
In (12),  $P_{ABL}(c_1|a_1\&b_k)$ is the probability given 
by the ABL rule using states $|a_1\rangle$ and 
$|b_k\rangle$ as inputs--just as in (7)--and
 $P_{\sigma_c}(b_k|a_1)$ is the final probability of 
outcome $b_k$
 given that the intervening $\sigma_c$ measurement was 
performed; i.e., the statistical weight of $\eta_k$.
 Thus $P_{\sigma_c}(b_k|a_1)$ is the analog of $P(b_k|a_1)$ in (8) 
for the case in which the intervening measurement has been 
performed. Explicitly, 
\vskip .2cm
$$P_{\sigma_c}(b_k|a_1) = P(b_k|c_1)P(c_1|a_1)+
P(b_k|c_2)P(c_2|a_1)\eqno(13)$$
\vskip .2cm
The right hand side of (13) is the weight of $\eta_k$, 
 the first term being the weight of subensemble $E'_{1k}$ 
and the second term being the weight of subensemble $E'_{2k}$. 
Using the rule of total probability and (13), it may easily
  be verified that equation (12) is identical to equation
 (10),  as follows.

	The right hand side of (12) can be rewritten,
 using (13), as
\vskip 1cm
\hskip 1cm $P_{ABL}(c_1|a_1\&b_1)\Bigl[ P(b_1|c_1)P(c_1|a_1)+
P(b_1|c_2)P(c_2|a_1)\Bigr]$ + \newline \smallskip  
\indent \hskip 3cm
 $P_{ABL}(c_1|a_1\&b_2)\Bigl[ P(b_2|c_1)P(c_1|a_1)+P(b_2|c_2)P(c_2|a_1)
\Bigr]$ \hskip 2.5cm (14)  
\vskip .2cm
	Noting that the quantities $P(b_k|c_j)$ are
 implicitly conditional also on $a_1$, i.e. they are 
equivalent to $P(b_k|c_j\&a_1)$, 
and for simplicity dropping the `ABL' subscript (which does
 not affect the calculation), (14) is equal to:
\vskip .2cm
\hskip 2cm $P(c_1|a_1\&b_1)\Bigl[ P(b_1\&c_1|a_1)+
P(b_1\&c_2|a_1)\Bigr] + \newline \indent
\hskip 4cm P(c_1|a_1\&b_2)
\Bigl[ P(b_2\&c_1|a_1)+P(b_2\&c_2|a_1)\Bigr]$ 
\hskip 4cm (15)\vskip .2cm
Using the rule of total probability, the quantities in
 brackets can be simplified, so that (15) is equal to
$$P(c_1|a_1\&b_1)\Bigl[ P(b_1|a_1)\Bigr] +P(c_1|a_1\&b_2)
\Bigl[ P(b_2|a_1)\Bigr]\eqno(16)$$
Expression (16) can be rewritten as:

$$P(c_1\&b_1|a_1) + P(c_1\&b_2|a_1)\eqno(17)$$
					
which, again applying the rule of total probability, 
yields finally

$$	P(c_1|a_1),\eqno(18)$$
which is exactly the quantity calculated in (10)
 using the conditional probability rule of
quantum mechanics. Therefore, despite the fact that the
statement of background condition $S_2'$
 does not explicitly appear in (9), the proposition $Q'$ 
derived from that
formula is the exact analog of $R$ for the case in which
$P$ holds; i.e.,  $Q'$ is equivalent to $R'$. 

	We have therefore established that $Q'$ as derived
 in the S\&S proof is indeed the
correct proposition to compare with $R (=Q$ of $(2))$. Note that 
$Q'$--which must be true--will, in
general,  differ from $R$  because the ABL rule referred to 
in $R (=Q)$ applies specifically 
to those systems in  ensembles  $E_i, i=1,2$,
  which actually obtain when the intervening
measurement has not been performed, i.e., when $P$
 does not hold ; and in general, it will
be the case that $E_i \ne \eta_k$ when i=k. Therefore, in 
general $R$ will be false.  (Further
clarification of the relationship between $Q, Q',  R$,
 and $R'$ will be made in the subsequent
discussion of Lewis' theory of counterfactuals and in 
the caption to Figure 3, which
illustrates points made in that discussion.) 

	Now we are in a position to pinpoint the error in 
Vaidman's counterargument II
as stated in his (1997a,b). The core of counterargument II
 is the claim that the S\&S proof
(as well as other inconsistency proofs along similar lines)
\footnote{For example, Cohen (1995) and Miller (1996).} is flawed 
in that it assumes
that no intervening measurement has occurred in the
 `counterfactual world' (which
corresponds to our hypothetical case of (6)). However,
 this claim is mistaken; it confuses
derivation  (5), performed to obtain the statement of 
actual background conditions $S_2$
entailed by the conjunction of $S_1 , L$ and $\neg P$,
 with the derivation applying to the
hypothetical case (6).

	It might be objected that $S= S_1 \& S_2$
 (in particular, $S_2$) is not the appropriate
statement of background conditions for deriving consequence 
$R$ in (6), and that we
should use instead $S'$ as in (6'). However, this  move 
cannot be justified for several reasons. 

	First,  it turns out that $S'$ is also not cotenable
 with $P$ in the given situation,
though this failure of cotenability is less obvious than
 that of $S$. The failure of
cotenability of $S'$ can best be seen through reference to
 Lewis' theory of counterfactuals
(1973, 1979). According to Lewis, a counterfactual is a
 `variably strict conditional': a
type of strict conditional requiring reference to what
 Lewis terms a set $\$_i$  of `spheres of
accessibility'  of possible worlds defined with respect 
to the actual world {\it i}.   Each sphere
$\zeta_i^{(k)} \in \$_i$  contains only those possible worlds
 that are similar  to the actual world {\it i}
according to a specified similarity relation for $\$_i$.
  The spheres are centered on {\it i} and
nested in a way that reflects their closeness, or degree
 of similarity, to {\it i}. In other words,  {\it i}
is at the center of $\$_i$  with the smallest sphere
 $\zeta_i^{(1)}$  comprising worlds that are more
similar to {\it i}  than are any of the worlds that are 
in spheres in  $\zeta_i^{(k)}$ , with  $k>1$, but not in 
$\zeta_i^{(1)}$ (And so on for $\zeta_i^{(2)}$, etc.).
\def\sphereset{$\$_i$}
\def\sphereik{$\zeta_i^{(k)}$}

	 According to Lewis, a counterfactual 
$P \ \cf Q$ is vacuously true at {\it i}  if and
only if no $P$-world (i.e., no possible world in which $P$
 is true) belongs to any sphere in $\$_i$, 
and non-vacuously true at {\it i}  if and only if some
 sphere $\zeta_i^{(k)}$  in \sphereset \ contains at least one
$P$-world  and the material conditional 
$P \to Q$  holds at every world in \sphereik
  (1973, p. 16). \sphereik  is also referred to
as a `P-permitting sphere.'  As noted earlier, these 
truth-conditions can be shown to be
fully equivalent to those in metalinguistic theories 
through reference to a natural
definition of cotenability in terms of Lewis' similarity
 relation between worlds. Thus,
Lewis defines a statement $X$ as cotenable with $P$
 at a world {\it i} if and only if either (1) $X$
holds throughout \sphereset \ or (2)  $X$ holds throughout
 some $P$-permitting sphere in \sphereset.  The truth 
conditions can now be stated in terms of cotenability: 
 $P\ \cf Q$ is true at {\it i}  iff $P$ and
some auxiliary premise $X$, cotenable with $P$ at {\it i},
 logically imply $Q$ (Lewis 1973, p. 57).

	Clearly, the statement $S =  S_1 \& S_2$
  is not cotenable with $P$ at {\it i} according to the
above definition. But neither is $S' = S_1 \& S_2'$ ,
 since it does not hold throughout any
$P$-permitting sphere in \sphereset   (see Figure 3).
\vskip .1cm
\hrule\input{fig3}
\nopagebreak
\def\aw{{\it i}}
\vskip 1pt\small \noindent \nopagebreak
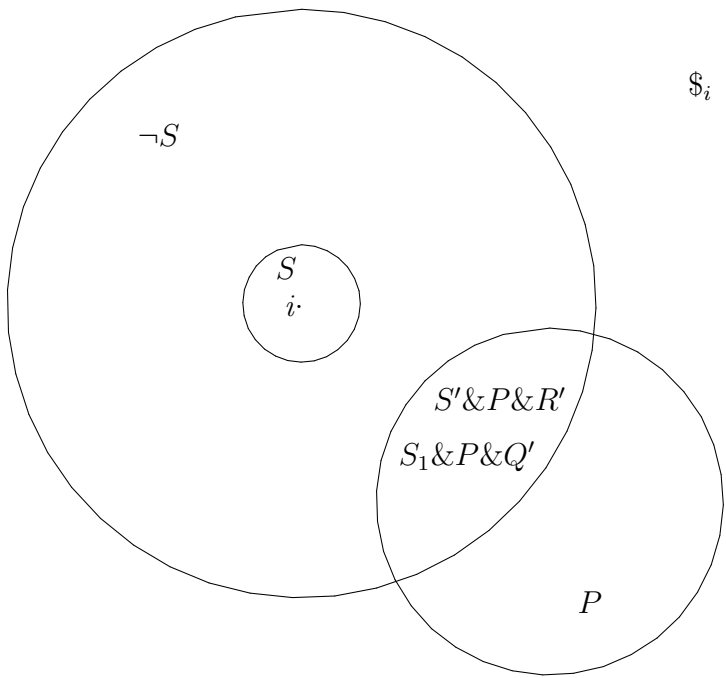
Figure 3. Neither $S$ nor $S'$ hold 
throughout any $P$-permitting sphere  in \sphereset.
(It is assumed that $L$ is true throughout \sphereset.) 
Note also that although we
have $S' \& P \to R'$ in the overlap region, and equivalently 
$S_1 \& P \to Q'$, we would need to have $P \to Q$ throughout a 
$P$-permitting sphere in order for the counterfactual 
$P \ \cf Q$ to be true. But in general, 
$Q' \ne Q$. The quantity $Q'=R'$ in the overlap region merely gives 
the ordinary quantum 
mechanical probability for a measurement that has actually
 been performed. Note that neither $R$ nor $Q$,
 being false, 
hold anywhere in \sphereset. (In exceptional cases where
it turns out that $S = S'$, we would have $Q'=R'=R=Q$,
so the counterfactual would be true. Such cases are discussed in
 Section 4.)
\hrule \vskip .2cm
\large
Therefore, the counterfactual (2) is still false if
(6) is modified by substituting $S'$ for $S$. Though such a
 substitution would allow us to
obtain the desired result that $R=Q'$, this statement would 
now apply not to a
counterfactual conditional but to a material conditional:
 the situation in which the
measurement has been made in the {\it actual} world. This is in
 fact what Vaidman proposes
in counterargument II, and it fails because, as noted 
earlier, it removes the counterfactual
element from the S\&S example.

 There is a more intuitive reason, however, why we should 
insist that the background
conditions $S = S_1 \& S_2$ are the correct ones for the 
counterfactual derivation (6) in this
particular case. This is  because the counterfactual 
reading of the ABL rule uses as input
quantities on record at time  $t_3 \ge t_2$. That is, the 
situation is one in which a pre- and
post-selection measurement (but no intervening measurement)
 have already been
performed; thus the experimenters are  in possession of 
recorded results for measurement
outcomes at both times $t_1$ and $t_2$. At that point, 
the question asked is a retrodictive
question: ``What would have happened had I measured 
observable $O$ at time $t, t_1 < t < t_2$?''
 In view of the existence of actual results at $t_2$,
 such results are an indelible part of the
history of world {\it i} and cannot be disregarded. Note, 
furthermore, that such  results are not
limited to single-system measurement outcomes but could also
 be statistical data on a
large number of measured systems, data which would clearly
 reflect the existence of ensembles $E_i$.
\vskip .3cm
{\bf  2. Time-symmetrized counterfactuals.~~}
\vskip .3cm

	In this section, I review and criticize two 
definitions given by Vaidman for a
proposed `time-symmetrized counterfactual,' which he claims
 is the only type of
counterfactual for which the ABL rule can be considered to
 give correct counterfactual
probabilities. I show that both definitions fail, in general,
  to establish a valid
counterfactual use of the ABL rule.  (In any case, it should
 be noted from the outset that
neither Goodman's, Stalnaker's nor Lewis'  theories are 
restricted to time-asymmetric
counterfactuals, i.e., counterfactuals whose consequent $Q$
 concerns events occurring at
times later than, or whose background conditions are limited
 to  events occurring at times
earlier than, that of events that are the subject of the 
antecedent $P$.)

	Vaidman (1996a, 1997a,b) argues that discrepancies 
like that in the S\&S proof
arise because they make use of a time-asymmetric
 counterfactual, which he says is
inappropriate for calculations involving the ABL rule. 
 He proposes a new type of
time-symmetrized counterfactual which he claims resolves 
these discrepancies. Vaidman
has provided two definitions. The first, which I will call
 Definition 1, appears as `interpretation c' in (1996a):
\vskip .3cm
\hskip 1.3cm \vbox{\hbox{Definition 1. ``Counterfactual probability [is] the
 probability for the results}\hbox{of a measurement if it 
has been performed in the world `closest' to the}
\hbox{actual world.'' (1996, p.9)}}
\vskip .2cm

 	Definition 1 is not  new,  since the idea of 
formulating truth-conditions for
counterfactuals in terms of a `closest possible world'
 has previously been proposed by
Stalnaker (1968).  The novelty of  Definition 1 as proposed 
by Vaidman consists  in   the   
choice of `selection function'  (in  Stalnaker's 
 terminology;  cf. 1968, p. 120)  that   determines the
 closest possible world. 
\def\cpw{{\it j}\ } \def\aw{{\it i}}The closest possible 
world \cpw is defined by Vaidman 
as the one in which all measurements, excluding the 
intermediate measurement asserted
by $P$, have the same outcomes as in \aw.  This amounts to
 ignoring the inequivalence of the
mixtures M and \mprime which gives rise to the
 cotenability problem discussed in section 2.

	However,  Vaidman's argument for the legitimacy of 
ignoring the difference between the mixture M in world \aw\ 
 and  the mixture \mprime in world \cpw is flawed.  His claim is 
that, since it is postulated that the system is not disturbed
 between $t$ and $t_2$--the period during which the mixtures
 differ--the difference between the mixtures has ``no
 physical meaning.'' (1996, p.10)  However, this is not
 the case, as can be shown if one considers an experiment 
along the lines of the S\&S example in which a large number
 N of instances is considered. In fact, I will now show
 that, due to the physical constraints imposed by quantum 
mechanics,  in general there can be, for large N, no such ``closest 
possible world'' \cpw . 
\def\aone{$|a_1\rangle$}
\def\bone{$|b_1\rangle$}
\def\btwo{$|b_2\rangle$}

	The requirement for  \cpw  is that
 all measurements,
 except the intervening one not occurring in \aw, have the
 same outcomes as in \aw.  Consider each run of the
 experiment in which the spin-{$1\over 2$} particle is
 preselected as `up along a' (\aone).  For each of
 these runs, when we find final outcome \bone in \aw \ we must
 also find final outcome \bone in \cpw and likewise for final
 outcome \btwo. 
\def\qab{\theta_{ab}}
\def\qac{\theta_{ac}}
\def\qcb{\theta_{cb}}

	The experiment in \aw \ consists of the initial 
 spin measurement along {\bf a} and one subsequent spin
measurement at an angle $\theta_{ab}$ (between directions 
{\bf a} and {\bf b}), while in \cpw  it consists of the
 initial {\bf a} spin measurement and two subsequent 
 measurements at relative angles $\theta_{ac}$ and
 $\theta_{cb}$. 

	So (assuming the spin directions are in the same 
plane) we must have 

$$\theta_{ab}= \theta_{ac} + \theta_{cb}.\eqno(19)$$

 Now,  the fraction $F_i(b_1)$ of runs resulting in the
 particle being found up along $\qab$ in \aw \ is given by the 
weight of ensemble $E_1$, i.e.:
\def\cosab{$cos^2({\theta_{ab}}\over 2$)}

$$F_i(b_1) = |\langle b_1|a_1\rangle|^2 = 
cos^2 {\left({\theta_{ab}} \over 2 \right)} \eqno(20)$$

 The fraction $F_j(b_1)$ of runs resulting in the particle 
being found up along $\qab$ in world \cpw is given by the 
weight of ensemble $\eta_1$ , i.e.:

$$F_j(b_1) =  |\langle b_1|c_1\rangle|^2
 |\langle c_1|a_1\rangle|^2 +
 |\langle b_1|c_2\rangle|^2 |\langle c_2|a_1\rangle|^2=
cos^2 {\left({\theta_{cb}} \over  2 \right)}
cos^2 {\left({\theta_{ac}} \over  2 \right)}+
sin^2 {\left({\theta_{cb}} \over  2 \right)}
sin^2 {\left({\theta_{ac}} \over  2 \right)}\eqno(21)$$

Since it is postulated that each run 
in \cpw has the same outcome as in \aw, we must  have
 $F_i(b_1) = F_j(b_1)$, i.e.:

$$cos^2 {\left({\theta_{ab}} \over  2 \right)} =
 cos^2 {\left({\theta_{cb}} \over  2 \right)}
cos^2 {\left({\theta_{ac}} \over  2 \right)}+
sin^2 {\left({\theta_{cb}} \over  2 \right)}
sin^2 {\left({\theta_{ac}} \over  2 \right)}
\eqno(22)$$

\noindent (In other words, the weight of ensemble $E_1$ must be equal
 to the weight of ensemble $\eta_1$).\footnote{\normalsize Actually, (22)
 is a necessary but not sufficient condition for validity of 
the counterfactual
use of the ABL rule. In certain cases, one may have the
 weights of $E_i$ and $\eta_k$  equal for
$i=k$, but the actual ensembles themselves cannot necessarily
 be identified: $E_i \ne \eta_k$ (c.f.
Bub and Brown, 1986, p. 2339). In showing, however, that (22)
 is not satisfied in general,
we show that even this necessary condition does not hold and 
therefore this definition for
a time-symmetrized counterfactual is, in general, untenable.}
 However,
 (22) is in general not satisfiable for $\qab= \qac
 + \qcb$.  For example, if we take  $\qac= \qcb = \pi/4$, 
we obtain:
$$1/2 \sim .728 + .02  \sim .749\eqno(23)$$
Since the requirement (22) cannot in general be fulfilled, there is 
in general no closest possible world \cpw to support  the symmetrized 
counterfactual that Vaidman proposes.

	 Vaidman has proposed a second definition for the time-symmetrized
counterfactual (1997a):
\vskip .3cm
\hskip 1.3cm \vbox{\hbox{Definition 2. ``If a measurement of
 an observable $C$ were
 performed at time $t$,}\hbox{then the probability for $C=c_i$
 would equal $p_i$, provided that the results of}
\hbox{measurements performed on the system at times 
$t_1$ and $t_2$ are fixed.'' (1996a, p.9)}}
\vskip .2cm

Definition 2 gives predictions consistent with quantum
 mechanics, if  the phrase ``the
results of measurements performed on the system at times 
$t_1$ and $t_2$ are fixed,''  is
intended to stipulate that the ABL rule only applies to those
 individual systems that (i) are pre- and
post-selected in the time-symmetrized state 
$\langle b_k||a_1\rangle, k=1,2$,  in the experiment actually
performed (in the S\&S example, no intervening measurement
 having been made) and (ii) {\it
would have been} pre- and post-selected in the {\it same} state if, 
counterfactually, a different
measurement had been performed (in the S\&S example, an 
intervening measurement having been made).  Since such a 
system is stipulated to have
the appropriate two-state vector for the counterfactual 
measurement as well as for the
actual measurement (in this case a measurement of the
 identity $I$), the non-counterfactual
reading of the ABL rule would in fact be applicable to such 
a system. 

	To clarify this point, let us augment the two-state 
vector notation with a subscript
specifying which observable has been measured at time $t$ in 
the pre- and post-selection of
system $K$. Thus, if system $K$ is preselected in state
\aone and post-selected in state $|b_k\rangle$ via
a measurement of observable $I$ at time $t$, the
 two-state vector $\Psi$ of system $K$ is 
$\Psi = \langle b_k||a_1\rangle_I$. Let us say that system
$K$ is "time-symmetrically fixed" in the sense of
Definition 2 iff we can
substitute a proposed counterfactually measured observable 
$O$ for the observable $I$ that
was actually measured, and $K$ would still have the
 same two-state vector. Thus we 
can also assign to $K$ the two-state vector
 $\Psi' = \langle b_k||a_1\rangle_O$. In virtue of the fact 
that K can
be described by $\Psi'$,  we can then employ the 
non-counterfactual reading of the ABL rule
to calculate the probability $P(x | a,b; O)$ where $x$
 is in the range of $O$, even though $O$ is
not the observable that was actually measured.

However, since we only exist and conduct experiments in one
 world, we cannot get to
other counterfactual worlds and perform concurrent 
experiments there to 
find which systems do have the same post-selection outcomes
 for various intervening
measurements. Since the counterfactual ABL probabilities only
 apply to the
time-symmetrically fixed systems, and we have no way of
 identifying these, any
information provided by a counterfactual interpretation of 
the ABL rule under Definition
2 would seem to have no clear physical meaning.

	An illustration of this difficulty of Definition 2 
is given in figure 4. In the actual
world \aw, we conduct an experiment in which electrons are 
pre- and post-selected in the
state $|z+\rangle$ (up along {\bf z}). We  perform an 
intervening measurement of $\sigma_{\theta}$, where 
$\theta={\pi \over 2}$.  If we start with
16 pre-selected electrons, a possible (ideal) distribution 
of measurement results is shown. 
\newpage 
\small
\input{fig4aa}\vskip .1cm
\input{fig4bb} \vskip .1cm \noindent
(a) In the actual world \aw, 16 systems preselected in the 
state $\vert z_1\rangle$ are subjected to an intervening
measurement of $\sigma_\theta$, where $\theta = {\pi \over 2}$
with respect to the z axis, and a final post-selection
measurement of $\sigma_z$. A possible (statistically ideal)
distribution is shown.\newline \noindent
(b)In a counterfactual world \cpw, the same 16
systems are subjected to an intervening measurement of
$\sigma_\phi$, where $\phi = {\pi \over 3}$ with respect
to the z axis, and again a final post-selection $\sigma_z$
measurement. A possible (statistically ideal) distribution
is shown. \newline\indent
Only the three systems in bold-face, $S_3,S_7$, and $S_{10}$, 
qualify as having their pre- and post-selection measurement 
outcomes `time-symmetrically fixed' for state 
$\langle z_1||z_1\rangle$. Applying the ABL rule for a
 counterfactual measurement of intervening observable 
$\sigma_\phi$ to any other systems having the symmetrized 
state $\langle z_1||z_1\rangle$ in world \aw \ would give 
incorrect probabilities for those systems. 
\hrule \large 

	Now, we ask the counterfactual question: 
``Consider a different angle $\phi = {\pi \over 3}$. 
 What would have been the probability of getting the result
 +1 were we to have measured $\sigma_\phi$ at $t$,
 instead of $\sigma_\theta$?'' 
In order to apply the ABL rule only to time-symmetrically
 fixed systems, we have to imagine this measurement 
performed
 concurrently in a counterfactual world \cpw, and choose only 
those systems post-selected in state $|z+\rangle$ in \cpw.
However,
 not all of the systems post-selected in the actual 
world \aw \ via the $\sigma_\theta$ measurement would 
necessarily be post-selected in \cpw 
via the $\sigma_\phi$  measurement. (See figure 4). 

	The only systems which are pre- and post-selected 
as $|z+\rangle$ in both worlds in this example are 
$S_3, S_7$, and $S_{10}$.
 These qualify as having their outcomes fixed in a
 time-symmetric sense. But since we are only actually 
performing the experiment in \aw, we do not know which of our 
8 systems with state $\langle z+||z+\rangle$ in \aw \ have
 fixed outcomes for
 other possible experiments in other possible worlds such 
as \cpw. For example, were we to apply the counterfactual ABL
 reading to systems $S_2, S_5, S_{12}, S_{15}$,
  or $S_{16}$,
 we would be using it incorrectly since these systems do 
not have outcome $|z+\rangle$ in the counterfactual world 
\cpw and 
therefore do not qualify as having fixed outcomes at 
both times, as is required for the symmetrized 
counterfactual. But there is no way to know this. 
Again, there is no way to make use of the information
 provided by the counterfactual ABL rule according to 
Definition 2; nor is it even clear that the notion of
 `time-symmetrically fixed' is physically meaningful.
\newpage

{\bf 4. Limited counterfactual ABL interpretation and the 
`advanced action' concept.~~ }	

	I will now discuss a special case in which Vaidman's
 time-symmetrized counterfactual interpretation may be 
correctly applied to the ABL rule. This is the case in which
 the hypothetical intervening measurement is of either the 
pre-selection observable $A$ or the post-selection 
observable $B$. Before showing how this special case holds 
for Vaidman's time-symmetrized counterfactual, I note that 
it falls under a class of counterfactuals identified by 
Cohen (1995) and Cohen and Hiley (1996) which  they 
obtained by applying the consistency conditions 
of Griffiths (1984) in his consistent histories 
interpretation.\footnote{\normalsize See also Griffiths
(1996) and (1998).}

	Cohen and Hiley found that a counterfactual 
interpretation of the ABL rule can be justified for a
 system preselected in state $|\psi_1(t_1)\rangle$
 and postselected in $|\psi_2(t_2)\rangle$  if, for every 
pair $\alpha , \beta$  of eigenvalues of 
operator $C$,

$$Re \{\langle \psi_2(t)|P_{\alpha} |\psi_1(t)\rangle
\langle\psi_1(t)|P_{\beta}|\psi_2(t)\rangle\} = 0.\eqno(24)$$
where $P_{\nu}$  is the projection operator onto the 
space of eigenstates with eigenvalue $\nu$  which are the 
possible outcomes of a measurement of $C$ at time $t$, 
$t_1<t<t_2$.\footnote{\normalsize Equation (24) is derived
under the assumption that the 
$P_{\nu}$ are components of a decomposition of the identity
operator in terms of orthogonal projections satisfying
the condition $P_{\lambda}P_{\nu} = \delta_{\lambda \nu} 
P_{\nu}$.}  It can easily be seen that the above special
 case (i.e., with $(\alpha , \beta)$
 from the set of eigenvalues either of observable $A$
 or observable $B$--not both together) satisfies (24). 

	For example, if we choose $B$ as the counterfactually
  measured observable, and the pre- and post-selected states 
as $|a_1\rangle$ and $|b_1\rangle$ respectively, with $H=0$, (24) 
becomes

$$Re \{\langle b_1|b_1\rangle\langle b_1|a_1\rangle
\langle a_1|b_2\rangle\langle b_2|b_1\rangle\} = 0.\eqno(24a)$$

	We see from the form of (24a) that any 
observable\footnote{\normalsize More precisely, any observable
satisfying the condition of footnote 9.} 
commuting with either the pre- or post-selection observable 
will give rise to at least one factor consisting of an inner 
product of orthogonal eigenvectors. Thus for such a case,
 (24) will always be satisfied.

	Condition (24) is a powerful one because it provides
 a basis for assertions about the ontological properties of 
systems between measurements: pre- and post-selection 
experimental histories satisfying (24) permit inferences 
about the states of systems even at times when no
 measurement is made.\footnote{\normalsize For a detailed discussion of
this interesting property, see Cohen (1995, p.4337).}

	Let us now confirm that both of Vaidman's definitions 
for a time-symmetrized 
counterfactual succeed for this case (as indeed they must, 
since even an ordinary time-asymmetric counterfactual reading 
of the ABL rule succeeds here).
 First, the existence of a closest possible world \cpw as 
postulated by Vaidman is permitted by  condition
 (22) in this case. If the observable  measured at time
$t$ is the same as either the pre- or post-selection 
observable, we  have either (i) $\qac=0$ or 
(ii) $\qcb=0$.  Putting each of these cases into (22),
 we obtain:

$$cos^2{({\qab\over 2})} = cos^2{({\qcb\over 2})}\eqno(25i)$$	
$$cos^2{({\qab\over 2})} = cos^2{({\qac\over 2})}\eqno(25ii)$$

By way of (19), we have also that 

\hskip 6cm $\qab = \qcb$ for $\qac = 0$, and\hskip 4.4cm (26i)
\vskip .2cm
\hskip 6cm $\qab = \qac$ for $\qcb = 0$;\hskip 5cm (26ii)

\noindent thus, (25i and ii) are correct in each case.  So therefore 
there does exist a `closest possible world' \cpw  for this 
special case, and Definition 1 is tenable.\footnote{\normalsize 
As an illustration of the point made in footnote 7, note that
 there is a case in which (22) is satisfied but (24) fails: 
the case in which  the intervening measurement is of a spin
 direction orthogonal to {\it both} {\bf a} and {\bf b}.
 (In this case, the spin directions are not coplanar.) This
 is a case in which the weights of $E_i$ and $\eta_k$
 are equal for $i=k$, but the ensembles themselves are not.}

	In addition, Definition 2 will also be 
tenable for this case, 
since all appropriately pre-selected systems which are
 post-selected via no intervening measurement would also, 
with probability 1, be post-selected via an intervening 
measurement of either the pre- or post-selection observable. 
Thus all pre- and post-selected systems in world \aw \ 
can be considered to be time-symmetrically fixed. We can see 
that this is the case by referring again to condition (24). 
Consider a system preselected in state $|a_1\rangle$
 and post-selected in state $|b_1\rangle$ via no intervening 
measurement. According to condition (24), were we to make an
 intervening measurement of observable $B$ at time $t$,
 the probability of outcome $|b_1\rangle$ would be correctly 
given by the ABL rule; and that probability would be 1. 
Given that the system is not disturbed between times $t$
 and $t_2$, a post-selection measurement of $B$
 must therefore find the system in state $|b_1\rangle$.
 So we can therefore say that if $B$ had been measured at 
time $t$, the same individual system would still have been 
post-selected.  

	In order to show that a pre- and post-selected 
system in state $\langle b_1||a_1\rangle$ via no intervening 
measurement would still be pre- and post-selected via an 
intervening measurement of observable $A$, we note that a 
counterfactual measurement of $A$ at a time $t_a, 
t_1<t_a<t_2$,  would trivially confirm that the system is in
state $|a_1\rangle$. Then we again have to use (24) together
with the ABL rule and a counterfactual $B$ measurement at
time $t_b,\newline t_1 < t_a < t_b < t_2,$ to show that the system
can be considered to be in the state $|b_1\rangle$
at a time $t_b$ after $t_a$ and before $t_2$, even though 
$B$ has not actually been measured.  If the system is in 
state $|b_1\rangle$ prior to $t_2$, and is not disturbed up 
until $t_2$, then it must be post-selected in that state
upon being subjected to a post-selection measurement of $B$
at $t_2$. Therefore, in this case, the system 
qualifies as time-symmetrically fixed, and Definition 2 is 
tenable. 

	Additional justification for a limited ABL 
counterfactual interpretation, i.e. for the above special case,
comes from Price's concept of `advanced action'. This notion 
will be briefly described below, along with his arguments
for the claim that we live in a world which exhibits
advanced action.
\def\microin{{$\mu$}innocence}

	Price (1996) raises an objection to the usual
interpretations of quantum mechanics based on their
explicit  time-asymmetry, which  assumes his contested
\microin \ (micro-innocence) assumption: the notion
that a quantum system's state depends on past interactions 
but not future ones. The objection consists in the claim, 
for which Price argues in his book, that there is no 
objective physical basis for such an asymmetry. 

	 Price (p.187) discusses an example of a photon
passing through two polarizers (a typical TSQT situation). 
He observes that our ordinary time-asymmetric intuition 
is to say that, with the history of the photon prior to 
the first polarizer held fixed (call this case (i)), the state of the photon 
between the first and second polarizers will not change 
if the orientation of the future polarizer is changed. 
In terms of ensembles, this is the usual quantum mechanical
way of defining states and ensembles in terms of a 
preselection measurement only. Now, in order to demonstrate 
the time-asymmetry of this intuition, Price asks us to 
consider the temporal mirror image, call it case (ii), 
of the above case, call it (i) (p. 187). Case (ii) is as follows: 
with the future of the photon after it passes the second 
polarizer held fixed, changes in the orientation of the 
first polarizer will not affect the state of the photon 
between the first and second polarizers. But we clearly 
reject this view, because we consider that the state of 
the photon after it passes the polarizer is dependent on 
the setting of that polarizer (p. 188).

	Price argues that because standard quantum mechanics 
accepts case (i) but rejects case (ii), it implicitly 
endorses an additional time asymmetry above and beyond 
what can be accounted for by the usual time-asymmetric 
convention for assessing counterfactuals (which assumes 
that only the past is fixed). Since the physics of 
the microworld reflects no such asymmetry, Price views 
the above time-asymmetric residue as unacceptable, and 
therefore argues that the world must exhibit advanced action.
If advanced action is admitted, then the asymmetry 
disappears because one rejects both cases above: the 
state of the photon in the region between the polarizers 
depends on the orientation of {\it both} the past and 
future polarizers. On such a view, it does not make sense to define 
an ensemble only in terms of a preselection measurement, 
because that single measurement does not yield the maximal 
information about the photon. In a world with advanced 
action, the photon has some additional predisposition based 
on the measurement to be made in its immediate future. 

In exactly the same way, TSQT advocates argue that the 
full description of a quantum system, both in terms of its 
state description and any ensemble of which it would be considered a 
member, requires that one take into account such future 
measurements. Hence they argue that such systems should be 
labeled by a two-state vector rather than a one-state vector 
(Price himself does not, however, refer to two-state 
vectors $\langle\Psi\Vert\Phi\rangle$
 in his discussion of the time-symmetry of quantum 
theory).

	How can the above considerations support a 
counterfactual reading of the ABL rule for the special case 
described above? In a world with advanced action, the 
causal effects of measurements are fully time-symmetric. 
That means we should be able to consider an experiment of 
the kind discussed above (i.e., the photon passing two 
polarizers) from either the usual temporal direction or the 
time-reversed direction, and either way should correspond 
to a valid physical process.

	Consider an experiment in which no observable
(equivalently, the indentity $I$) is
measured at the intermediate time $t$. Let our photon be 
described by the two-state vector $\langle b_1||a_1\rangle$;
i.e., it will be pre- and post-selected to begin with 
eigenvalue $a_1$ and to end up with eigenvalue $b_1$. 
Let the ABL probability of outcome $x$ at time $t$
be denoted by $P(x;t)$. Under the assumption that 
the ABL rule is valid {\it only} for intervening measurements 
that have actually been performed (call this the `strict 
ABL assumption'), it is invalid to use the ABL 
rule to calculate $P(b_1;t)$ for a counterfactual measurement
of $B$ in this situation. Yet, we also know that 
the photon is going to end up in state $|b_1\rangle$
at time $t_2$.  The ABL rule gives the value 1 for 
$P(b_1;t)$. If the system is not disturbed at all between 
$t$ and $t_2$ (as is assumed), then this result seems 
fully consistent with that fact. 
\def\prevx{$P\!\raisebox{2 pt}{*}\!(x;t)$}
\def\prevbone{$P\!\raisebox{2 pt}{*}\!(b_1;t)$}
\def\prevaone{$P\!\raisebox{2 pt}{*}\!(a_1;t)$}

	Now, one might argue that this merely reflects
the arbitrary decision to postselect for the state 
$|b_1\rangle$ at $t_2$, and that the value of $P(b_1;t)$ 
has nothing to do with the ontological state of the 
photon at that time (or at any time prior to $t_2$). 
But this argument is weakened in a fully time-symmetric 
world with advanced action. This can be seen by viewing 
the same process from the time-reversed direction. 
(In the following, we denote ABL probabilities applying 
to the time-reversed case by \prevx.)
 
	Suppose we have a system ``starting out'' 
at time $t_2$ in state $|b_1\rangle$, no measurement
occurring at time $t$, and the system being found in 
state $|a_1\rangle$ at time $t_1$. Were we to 
calculate \prevbone and get the value 1, according to the 
strict ABL assumption we would be using the ABL rule 
incorrectly. Yet the result \prevbone = 1 correctly 
describes the actual physical situation: we know that the 
photon must have the eigenvalue $b_1$ at time $t$, 
since it ``started out'' with that value (viewing the 
post-selection at $t_2$ as the starting point). 

	The fact that we view the ``incorrect'' 
ABL usage in a different light depending on the temporal 
direction considered should alert us to the fact that, 
as Vaidman or Price would put it, we are operating under a 
``time asymmetry prejudice.'' If we truly want to rule out 
the ``incorrect'' use of the ABL rule, such a judgment 
should stand up to scrutiny when viewed from either 
temporal direction. Yet, as shown above, when viewed in 
the time-reversed direction, our argument for the rejection 
of the ``incorrect'' usage is considerably weaker. (A
similar argument, though not explicitly in terms of
the ABL rule, can be found in Griffiths (1984, pp. 238-9).)	

{\bf  5. Conclusion.~~}

	It has been argued that the counterfactual 
interpretation of the ABL rule is not valid in general. 
For the usual  time-asymmetric counterfactual situation, 
as in the Sharp and Shanks proof, it gives results generally 
inconsistent with quantum mechanics. Counterarguments by 
Vaidman (1996a, 1997a,b) against this proof and against others like 
it (Cohen, 1995 and Miller, 1996) fail  because they fail 
to take into account the fundamental cotenability problem 
posed by the counterfactual use of the ABL rule, and 
mistakenly identify as a flaw what is actually an 
appropriate application of  background conditions for 
evaluation of  the counterfactual statement.  Proposed 
definitions for a new kind of time-symmetric counterfactual 
are problematic in that they rely either on the notion of a 
`closest possible world' that in general does not exist, 
or on a time-symmetric fixing requirement that is generally 
impossible to fulfill. 

	However, there is an interesting special case in 
which the ABL rule may be correctly used in a counterfactual 
sense: that in which the hypothetical intervening 
measurement is that of an operator that commutes with 
either the pre- or post-selection operator.\footnote 
{\normalsize This particular case is addressed 
in this paper because advocates of the counterfactual
interpretation of the ABL rule have cited it as an example
of ``curious properties'' of pre- and post-selected ensembles.} 
This case is 
a member of a class identified by Cohen and Hiley which 
satisfies a condition derived from Griffith's 
consistent histories approach. In this case, both 
definitions of the time-symmetrized counterfactual are 
satisfied: there does exist a closest possible world, 
and the time-symmetrical fixing requirement can be 
fulfilled. Moreover, a counterfactual interpretation of 
the ABL probabilities in this case can be further supported 
by considering a physical picture that includes Price's 
advanced action concept.  

{\bf  Acknowledgments~~}

	I am grateful to my dissertation advisor, Jeffrey 
Bub, for introducing me to this topic and for valuable 
discussions. I have also benefited from discussions and 
correspondence with Robert Griffiths, 
Robert Ryna\nobreak siewicz, Abner Shimony and Lev Vaidman, and 
comments from Jeremy Butterfield and two anonymous referees.

{\bf  References~~}

\noindent Aharonov, Y, Bergmann, P.G. , and Lebowitz, J.L. (1964), 
`Time Symmetry in the Quantum Process of Measurement,' 
{\it Physical Review B 134}, 1410-16.\newline
Aharonov, Y. and Vaidman, L. (1991), `Complete Description
of a Quantum System at a Given Time,'
{\it Journal of Physics A 24}, 2315-28.\newline
Albert, D.Z., Aharonov, Y., and D'Amato, S. (1985), 
{\it Physical Review Letters 54}, 5.\newline
Bub, J. and Brown, H. (1986), `Curious Properties of 
Quantum Ensembles Which
Have Been Both Preselected and Post-Selected,'  
{\it Physical Review Letters 56}, 2337-2340.\newline
Cohen, O. (1995), `Pre- and postselected quantum systems, 
counterfactual 
measurements, and consistent histories, '
{\it Physical Review A 51}, 4373-4380.\newline
Cohen, O. and Hiley, B.J. (1996), `Elements of Reality, 
Lorentz Invariance, and the Product Rule,' 
{\it Foundations of Physics 26}, 1-15.\newline
Goodman, N. (1947), `The Problem of Counterfactual 
Conditionals,' {\it Journal of Philosophy 44}, 113-128.
\newline
Griffiths, R. B. (1984), `Consistent Histories
and the Interpretation of Quantum Mechanics,' 
{\it Journal of Statistical 
Physics 36}, 4373.\newline
Griffiths, R. B. (1996), ``Consistent Histories and Quantum Reasoning,'' 
{\it Phys. Rev. A 54}, 2759.\newline
Griffiths, R. B. (1998), ``Choice of consistent family, and quantum
incompatibility,'' {\it Phys. Rev. A 57}, 1604.\newline
Horwich, P. (1988), {\it Asymmetries in Time}, 
Cambridge: MIT Press.\newline
Lewis, D. (1973), {\it Counterfactuals}, Cambridge: 
Harvard University Press. \newline 
Lewis, D. (1979), `Counterfactual Dependence and Time's 
Arrow,' {\it Nous 13}, 455-476.\newline
Mermin, N. D. (1997), 'How to Ascertain the Values of 
Every Member of a Set of Observables That Cannot All 
Have Values,' in R. S. Cohen et al. (eds), 
{\it Potentiality, Entanglement and Passion-at-a-Distance}, 
149-157, Kluwer Academic Publishers.\newline
Miller, D.J. (1996), `Realism and Time Symmetry in 
Quantum Mechanics,' {\it Physics Letters A 222}, 31.\newline
Price, H. (1996), {\it Time's Arrow and Archimedes' Point},
Oxford University Press.\newline
Sharp, W. and Shanks, N. (1993), `The Rise and Fall of 
Time-Symmetrized 
Quantum Mechanics,' {\it Philosophy of Science 60}, 
488-499.\newline
Stalnaker, R. (1968), `A Theory of Conditionals,' 
in {\it Studies in Logical Theory}, edited by N. Rescher. 
Oxford: Blackwell.\newline
Vaidman, L. (1996a), `Defending Time-Symmetrized Quantum 
Theory,' Tel-Aviv University preprint, 
quant-phys/9609007. \newline
Vaidman, L. (1996b), `Weak-Measurement Elements of Reality,' 
{\it Foundations of Physics 26}, 895-906.\newline
Vaidman, L. (1997a), `Time-Symmetrized Counterfactuals 
in Quantum Theory,' Tel-Aviv University preprint 
TAUP 2459-97.\newline
Vaidman, L. (1997b), `Time-Symmetrized Quantum Theory,' 
invited lecture, Fundamental Problems in Quantum Theory 
workshop, University of Maryland Baltimore County, 
Aug. 4-7, 1997; quant-ph/9710036.

\end{document}

%% file: fig1.tex
\special{em:linewidth 0.4pt}
\unitlength 1.00mm
\linethickness{0.4pt}
\begin{picture}(136.00,100)
\put(84.67,115.67){\makebox(0,0)[rc]{Figure 1}}
\put(43.67,81.66){\makebox(0,0)[rc]{$|a_1\rangle$}}
\put(136.00,96.33){\makebox(0,0)[cc]
{$|b_1\rangle, E_1\!:\Psi_1=\!\langle b_1||a_1\rangle$}}
\put(47.67,20.00){\makebox(0,0)[cc]{$t_1$}}
\put(95.00,19.00){\makebox(0,0)[cc]{$t$}}
\put(114.33,19.00){\makebox(0,0)[cc]{$t_2$}}
\emline{47.67}{81.67}{1}{115.33}{96.33}{2}
\emline{115.33}{96.33}{3}{115.33}{96.33}{4}
\emline{48.00}{81.67}{5}{115.00}{63.67}{6}
\put(121.33,23.00){\vector(1,0){0.2}}
\emline{43.67}{23.00}{3}{121.33}{23.00}{4}
\put(135.67,63.00){\makebox(0,0)[cc]
{$|b_2\rangle, E_2\!:\Psi_2=\!\langle b_2||a_1\rangle$}}
\end{picture}

%% file: fig2.tex
\special{em:linewidth 0.4pt}
\unitlength 1mm
\linethickness{0.4pt}
\begin{picture}(110.00,100.67)(0,35)
\emline{25.67}{90.33}{1}{63.33}{101.67}{2}
\emline{26.33}{90.33}{3}{63.00}{79.33}{4}
\emline{74.00}{104.67}{5}{95.00}{110.67}{6}
\emline{74.33}{104.67}{7}{94.67}{97.33}{8}
\emline{73.67}{76.33}{9}{95.33}{81.33}{10}
\emline{74.33}{76.33}{11}{95.67}{68.67}{12}
\put(109.33,47.00){\vector(1,0){0.2}}
\emline{19.67}{47.00}{3}{109.33}{47.00}{4}
\put(61.00,118.67){\makebox(0,0)[cc]{Figure 2}}
\put(20.33,90.33){\makebox(0,0)[cc]{$|a_1\rangle$}}
\put(68.33,103.33){\makebox(0,0)[cc]{$|c_1\rangle$}}
\put(68.33,77.67){\makebox(0,0)[cc]{$|c_2\rangle$}}
\put(120.67,111.33){\makebox(0,0)[cc]
{$|b_1\rangle,E'_{11}\!:\Psi'_{11}=\langle b_1||a_1\rangle^{(c_1)}$}}
\put(120.67,96.67){\makebox(0,0)[cc]
{$|b_2\rangle,E'_{12}\!:\Psi'_{12}={\langle b_2||a_1\rangle}^{(c_1)}$}}
\put(120.67,81.67){\makebox(0,0)[cc]
{$|b_1\rangle,E'_{21}\!:\Psi'_{21}=\langle b_1||a_1\rangle^{(c_2)}$}}
\put(120.67,67.67){\makebox(0,0)[cc]
{$|b_2\rangle,E'_{22}\!:\Psi'_{22}=\langle b_2||a_1\rangle^{(c_2)}$}}
\put(25.67,42.33){\makebox(0,0)[cc]{$t_1$}}
\put(70.00,42.67){\makebox(0,0)[cc]{$t$}}
\put(97.00,43.00){\makebox(0,0)[cc]{$t_2$}}
\end{picture}

%% file: fig3.tex
\special{em:linewidth 0.4pt}
\unitlength 1.00mm
\linethickness{0.4pt}
\begin{picture}(107.40,120.00)(0,37)
\emline{51.33}{105.48}{1}{53.07}{105.28}{2}
\emline{53.07}{105.28}{3}{54.73}{104.70}{4}
\emline{54.73}{104.70}{5}{56.21}{103.76}{6}
\emline{56.21}{103.76}{7}{57.45}{102.52}{8}
\emline{57.45}{102.52}{9}{58.38}{101.03}{10}
\emline{58.38}{101.03}{11}{58.95}{99.37}{12}
\emline{58.95}{99.37}{13}{59.14}{97.63}{14}
\emline{59.14}{97.63}{15}{58.93}{95.89}{16}
\emline{58.93}{95.89}{17}{58.34}{94.23}{18}
\emline{58.34}{94.23}{19}{57.40}{92.75}{20}
\emline{57.40}{92.75}{21}{56.15}{91.52}{22}
\emline{56.15}{91.52}{23}{54.65}{90.60}{24}
\emline{54.65}{90.60}{25}{52.99}{90.04}{26}
\emline{52.99}{90.04}{27}{51.25}{89.86}{28}
\emline{51.25}{89.86}{29}{49.50}{90.08}{30}
\emline{49.50}{90.08}{31}{47.86}{90.68}{32}
\emline{47.86}{90.68}{33}{46.38}{91.63}{34}
\emline{46.38}{91.63}{35}{45.16}{92.88}{36}
\emline{45.16}{92.88}{37}{44.24}{94.38}{38}
\emline{44.24}{94.38}{39}{43.69}{96.05}{40}
\emline{43.69}{96.05}{41}{43.52}{97.80}{42}
\emline{43.52}{97.80}{43}{43.75}{99.54}{44}
\emline{43.75}{99.54}{45}{44.35}{101.18}{46}
\emline{44.35}{101.18}{47}{45.31}{102.65}{48}
\emline{45.31}{102.65}{49}{46.58}{103.87}{50}
\emline{46.58}{103.87}{51}{48.08}{104.77}{52}
\emline{48.08}{104.77}{53}{51.33}{105.48}{54}
\emline{51.33}{136.81}{55}{56.95}{136.40}{56}
\emline{56.95}{136.40}{57}{62.46}{135.19}{58}
\emline{62.46}{135.19}{59}{67.74}{133.20}{60}
\emline{67.74}{133.20}{61}{72.68}{130.48}{62}
\emline{72.68}{130.48}{63}{77.17}{127.07}{64}
\emline{77.17}{127.07}{65}{81.13}{123.05}{66}
\emline{81.13}{123.05}{67}{84.46}{118.50}{68}
\emline{84.46}{118.50}{69}{87.11}{113.53}{70}
\emline{87.11}{113.53}{71}{89.02}{108.22}{72}
\emline{89.02}{108.22}{73}{90.15}{102.69}{74}
\emline{90.15}{102.69}{75}{90.47}{97.06}{76}
\emline{90.47}{97.06}{77}{89.97}{91.45}{78}
\emline{89.97}{91.45}{79}{88.68}{85.96}{80}
\emline{88.68}{85.96}{81}{86.61}{80.71}{82}
\emline{86.61}{80.71}{83}{83.80}{75.82}{84}
\emline{83.80}{75.82}{85}{80.32}{71.38}{86}
\emline{80.32}{71.38}{87}{76.25}{67.48}{88}
\emline{76.25}{67.48}{89}{71.65}{64.22}{90}
\emline{71.65}{64.22}{91}{66.63}{61.64}{92}
\emline{66.63}{61.64}{93}{61.29}{59.82}{94}
\emline{61.29}{59.82}{95}{55.75}{58.78}{96}
\emline{55.75}{58.78}{97}{50.12}{58.55}{98}
\emline{50.12}{58.55}{99}{44.51}{59.13}{100}
\emline{44.51}{59.13}{101}{39.04}{60.51}{102}
\emline{39.04}{60.51}{103}{33.83}{62.66}{104}
\emline{33.83}{62.66}{105}{28.98}{65.54}{106}
\emline{28.98}{65.54}{107}{24.59}{69.09}{108}
\emline{24.59}{69.09}{109}{20.76}{73.23}{110}
\emline{20.76}{73.23}{111}{17.57}{77.87}{112}
\emline{17.57}{77.87}{113}{15.07}{82.93}{114}
\emline{15.07}{82.93}{115}{13.33}{88.29}{116}
\emline{13.33}{88.29}{117}{12.38}{93.85}{118}
\emline{12.38}{93.85}{119}{12.23}{99.49}{120}
\emline{12.23}{99.49}{121}{12.90}{105.09}{122}
\emline{12.90}{105.09}{123}{14.36}{110.54}{124}
\emline{14.36}{110.54}{125}{16.60}{115.71}{126}
\emline{16.60}{115.71}{127}{19.55}{120.52}{128}
\emline{19.55}{120.52}{129}{23.16}{124.85}{130}
\emline{23.16}{124.85}{131}{27.36}{128.61}{132}
\emline{27.36}{128.61}{133}{32.06}{131.74}{134}
\emline{32.06}{131.74}{135}{37.15}{134.15}{136}
\emline{37.15}{134.15}{137}{42.54}{135.81}{138}
\emline{42.54}{135.81}{139}{51.33}{136.81}{140}
\emline{84.33}{94.40}{141}{88.39}{94.04}{142}
\emline{88.39}{94.04}{143}{92.32}{92.97}{144}
\emline{92.32}{92.97}{145}{96.00}{91.23}{146}
\emline{96.00}{91.23}{147}{99.32}{88.87}{148}
\emline{99.32}{88.87}{149}{102.17}{85.96}{150}
\emline{102.17}{85.96}{151}{104.47}{82.59}{152}
\emline{104.47}{82.59}{153}{106.13}{78.87}{154}
\emline{106.13}{78.87}{155}{107.12}{74.92}{156}
\emline{107.12}{74.92}{157}{107.40}{70.86}{158}
\emline{107.40}{70.86}{159}{106.95}{66.81}{160}
\emline{106.95}{66.81}{161}{105.80}{62.90}{162}
\emline{105.80}{62.90}{163}{103.99}{59.25}{164}
\emline{103.99}{59.25}{165}{101.55}{55.98}{166}
\emline{101.55}{55.98}{167}{98.59}{53.19}{168}
\emline{98.59}{53.19}{169}{95.17}{50.97}{170}
\emline{95.17}{50.97}{171}{91.42}{49.38}{172}
\emline{91.42}{49.38}{173}{87.45}{48.47}{174}
\emline{87.45}{48.47}{175}{83.38}{48.28}{176}
\emline{83.38}{48.28}{177}{79.34}{48.81}{178}
\emline{79.34}{48.81}{179}{75.46}{50.03}{180}
\emline{75.46}{50.03}{181}{71.85}{51.93}{182}
\emline{71.85}{51.93}{183}{68.63}{54.42}{184}
\emline{68.63}{54.42}{185}{65.90}{57.45}{186}
\emline{65.90}{57.45}{187}{63.75}{60.91}{188}
\emline{63.75}{60.91}{189}{62.24}{64.69}{190}
\emline{62.24}{64.69}{191}{61.41}{68.68}{192}
\emline{61.41}{68.68}{193}{61.30}{72.75}{194}
\emline{61.30}{72.75}{195}{61.91}{76.78}{196}
\emline{61.91}{76.78}{197}{63.22}{80.64}{198}
\emline{63.22}{80.64}{199}{65.19}{84.21}{200}
\emline{65.19}{84.21}{201}{67.75}{87.37}{202}
\emline{67.75}{87.37}{203}{70.83}{90.04}{204}
\emline{70.83}{90.04}{205}{74.33}{92.12}{206}
\emline{74.33}{92.12}{207}{78.15}{93.56}{208}
\emline{78.15}{93.56}{209}{84.33}{94.40}{210}
\put(49.67,97.33){\makebox(0,0)[cc]{{\it i}}}
\put(51,97.33){\makebox(0,0)[cc]{{.}}}
\put(49.33,102.33){\makebox(0,0)[cc]{$S$}}
\put(32.33,120.00){\makebox(0,0)[cc]{$\neg S$}}
\put(104.33,126.67){\makebox(0,0)[cc]{$\$_ i$}}
\put(59.33,145.00){\makebox(0,0)[cc]{Figure 3}}
\put(89.67,58.00){\makebox(0,0)[cc]{$P$}}
\put(77.67,85.00){\makebox(0,0)[cc]{$S'\& P\& R'$}}
\put(73.33,77.33){\makebox(0,0)[cc]{$S_1\& P\& Q'$}}
\end{picture}

%% file: fig4aa.tex
\special{em:linewidth 0.4pt}
\unitlength 1.00mm
\linethickness{0.4pt}
\begin{picture}(108.67,45)(0,85)
\emline{24.33}{105.00}{1}{63.00}{112.67}{2}
\emline{24.67}{105.00}{3}{63.00}{98.00}{4}
\emline{70.67}{114.33}{5}{83.67}{120.67}{6}
\emline{71.00}{114.00}{7}{83.67}{109.67}{8}
\emline{71.33}{96.33}{9}{84.33}{100.00}{10}
\emline{71.67}{96.33}{11}{84.33}{89.33}{12}
\put(50.67,150.00){\makebox(0,0)[cc]{\large Figure 4}}
\put(10.00,105.00){\makebox(0,0)[cc]{(a)}}
\put(19.67,105.00){\makebox(0,0)[cc]{$|z_1\rangle$}}
\put(67.00,113.67){\makebox(0,0)[cc]{$|\theta_1\rangle$}}
\put(67.33,96.00){\makebox(0,0)[cc]{$|\theta_2\rangle$}}
\put(87.00,121.33){\makebox(0,0)[cc]{$|z_1\rangle$}}
\put(87.00,109.00){\makebox(0,0)[cc]{$|z_2\rangle$}}
\put(87.67,99.67){\makebox(0,0)[cc]{$|z_1\rangle$}}
\put(87.67,88.00){\makebox(0,0)[cc]{$|z_2\rangle$}}
\put(96.66,122.00){\makebox(0,0)[lc]{$S_2, {\bf S_7}, {\bf S_{10}}, S_{12}$}}
\put(107.99,108.66){\makebox(0,0)[cc]{$S_1, S_4, S_6, S_8$}}
\put(108.67,99.33){\makebox(0,0)[cc]{${\bf S_3}, S_5, S_{15}, S_{16}$}}
\put(108.33,87.67){\makebox(0,0)[cc]{$S_9, S_{11}, S_{13}, S_{14}$}}
\end{picture}

%% file: fig4bb.tex
\special{em:linewidth 0.4pt}
\unitlength 1.00mm
\linethickness{0.4pt}
\begin{picture}(100.66,45)(0,85)
\emline{24.33}{105.00}{1}{63.00}{112.67}{2}
\emline{24.67}{105.00}{3}{63.00}{98.00}{4}
\emline{70.67}{114.33}{5}{83.67}{120.67}{6}
\emline{71.00}{114.00}{7}{83.67}{109.67}{8}
\emline{71.33}{96.33}{9}{84.33}{100.00}{10}
\emline{71.67}{96.33}{11}{84.33}{89.33}{12}
\normalsize
\put(10.00,105.00){\makebox(0,0)[cc]{(b)}}
\put(19.67,105.00){\makebox(0,0)[cc]{$|z_1\rangle$}}
\put(67.00,113.67){\makebox(0,0)[cc]{$|\phi_1\rangle$}}
\put(67.33,96.00){\makebox(0,0)[cc]{$|\phi_2\rangle$}}
\put(87.00,121.33){\makebox(0,0)[cc]{$|z_1\rangle$}}
\put(87.00,109.00){\makebox(0,0)[cc]{$|z_2\rangle$}}
\put(87.67,99.67){\makebox(0,0)[cc]{$|z_1\rangle$}}
\put(87.67,88.00){\makebox(0,0)[cc]{$|z_2\rangle$}}
\put(119.66,122.34){\makebox(0,0)[cc]
{$S_1, {\bf S_3}, S_4, S_6, {\bf S_7}, {\bf S_{10}}, S_{11}, S_{13}, S_{14}$}}
\put(101.33,109.33){\makebox(0,0)[cc]{$S_2, S_5, S_9$}}
\put(96.00,100.00){\makebox(0,0)[cc]{$S_8$}}
\put(102.99,87.34){\makebox(0,0)[cc]{$S_{12}, S_{15}, S_{16}$}}
\end{picture}